\documentclass{llncs}
\usepackage{hyperref}
\usepackage[numbers, comma, sort]{natbib}
\begin{document}

\title{Digital Repository of Mathematical Formulae\thanks{The final publication is available at \url{http://link.springer.com}.}\\[-0.1cm]}

%
\author{
Howard S. Cohl\inst{1}
\and
   Marjorie A.~McClain\inst{1}
\and
   Bonita V.~Saunders\inst{1}
\and 
   Moritz Schubotz\inst{2}
\and 
   Janelle C.~Williams \inst{3}
}

\institute{Applied and Computational Mathematics Division,
National Institute of Standards and Technology (NIST),
Gaithersburg, Maryland, U.S.A.,
\email{howard.cohl@nist.gov, marjorie.mcclain@nist.gov, bonita.saunders@nist.gov}
\and
Database Systems and Information Management Group,
Technische Universit\"{a}t, Berlin, Germany,
\email{schubotz@tu-berlin.de}\\
\and
Department of Mathematics and Computer Science,
Virginia State University, Petersburg, Virginia, U.S.A.,
\email{janelle.williams35@gmail.com}\\
}
\authorrunning{Cohl, McClain, Saunders, Schubotz and Williams} 


\clearpage

\maketitle

\vspace{-0.67cm}
\begin{abstract}
The purpose of the NIST Digital Repository of Mathematical Formulae (DRMF) is to create a digital compendium of mathematical formulae for orthogonal polynomials and special functions (OPSF) and of associated mathematical data. The DRMF addresses needs of working mathematicians, physicists and engineers: providing a platform for publication and interaction with OPSF formulae on the web.  Using \mbox{MediaWiki} extensions and other existing technology (such as software and macro collections developed for the NIST Digital Library of Mathematical Functions), the DRMF acts as an interactive web domain for OPSF formulae.  Whereas Wikipedia and other web authoring tools manifest notions or descriptions as first class objects, the DRMF does that with mathematical formulae. 
See \url{http://gw32.iu.xsede.org/index.php/Main_Page}.\\[-0.3cm]
\end{abstract}

\vspace{-1.20cm}
\section{Introduction}
\label{sect:introduction}
\vspace{-0.1cm}
Compendia of mathematical formulae have a long and rich history. Many
scientists have developed such repositories as books and these have been
extremely useful to scientists, mathematicians and engineers over the last
several centuries (see
\cite{
Brych,
ByrdFriedman,
ErdelyiHTF,
ErdelyiTIT,
Grad,
MOS,
PrudBrychMar}
for instance).
While there may be some overlap of formulae in different compendia,
one often needs to be familiar with many different compendia to find
a specific desired formula.
Online compendia of mathematical
formulae exist, such as the 
\href{http://dlmf.nist.gov}{NIST Digital Library of Mathematical Functions}
(DLMF), subsets of \href{http://en.wikipedia.org}{Wikipedia}, and
the \href{http://functions.wolfram.com}{Wolfram Functions Site}.
We hope to take advantage of the best aspects of these online efforts while 
also incorporating powerful new features that a community-arm of scientists 
should find beneficial.  
Our strategy is to start with validated and trustworthy special function data from the
NIST DLMF, while adding Web 2.0 capabilities which will encourage community members to 
discuss mathematical data associated with formulae.
These discussions will include internally hyperlinked proofs 
as well as mathematical connections between formulae in the repository.  

The online repository will be designed for a 
mathematically literate audience and should 
(1) facilitate interaction among a community of mathematicians and scientists 
interested in formulae data related to orthogonal polynomials and special functions
(OPSF);
(2) be expandable, allowing the input of new formulae;
(3) be accessible as a standalone resource;
(4) have a user friendly, consistent, and hyperlinkable viewpoint and authoring 
perspective; and
(5) contain easily searchable mathematics and take advantage of
modern MathML tools for easy to read, scalably rendered mathematics.
It is the desire of our group to build a tool that
brings the above features together in a public website for mathematicians,
scientists and engineers.
We refer to this web tool as the Digital Repository
of Mathematical Formulae (\href{http://drmf.mathweb.org}{DRMF}).

Our project was motivated by the goal of creating an interactive online compendium of 
mathematical formulae.  This need was was addressed in \href{https://www.siam.org}{SIAM}
Activity Group OPSF-Net discussions, such as 
\href{http://math.nist.gov/~DLozier/OPSFnet/OPSF-Net_18.4.pdf}
{Dmitry Karp (OPSF-Net 18.4, Topic \#5)}.
In that OPSF-Net edition, there were two related posts 
\href{http://math.nist.gov/~DLozier/OPSFnet/OPSF-Net_18.4.pdf}
{(OPSF-Net 18.4, Topics \#6,\#7)}
with a follow-up post in
\href{http://math.nist.gov/~DLozier/OPSFnet/OPSF-Net_18.6.pdf}
{OPSF-Net 18.6, Topic \#3}.  

\vspace{-0.3cm}
\section{Implementation}

\vspace{-0.2cm}
In our project, we have taken advantage of the free and open source \href{http://www.mediawiki.org}{MediaWiki} wiki software as well as
tools developed within the \href{http://dlmf.nist.gov}{DLMF project} \cite{NIST},
such as \LaTeX ML and the DLMF \LaTeX\,\,macros.
DLMF macros (and extensions as necessary) tie specific character sequences to 
unique mathematical objects such as 
special functions, orthogonal polynomials, or to other mathematical symbols 
associated with these.  The DLMF macros are hence used to define OPSF within
DRMF and through \LaTeX ML, their corresponding rendered mathematical symbols.  
Furthermore, the use of DLMF macros as linked to their definitions within the DLMF, 
allows for easy access to precise OPSF definitions for the symbols used within 
the \LaTeX\,\,source for OPSF formulae.  The committed use of DLMF macros 
guarantees a mathematical and structural consistency throughout the DRMF.
As a web tool, the DRMF provides
(1) formula interactivity,
(2) formula home pages, 
(3) centralized bibliography, 
and (4) mathematical search.
The DRMF shares the core DLMF component, 
\href{http://dlmf.nist.gov/LaTeXML}{\LaTeX ML}, which (through 
the \mbox{MediaWiki} math extension) processes Wikitext math 
markup written in \LaTeX\,\,to produce XML and HTML.
For formula interactivity and menus linked to formulae, we have utilized
the \href{https://github.com/KWARC/jobad}{JOBAD interactivity
framework} and are investigating the use of MathJax \cite{Cervone}.
We have also 
incorporated the \href{http://www.mediawiki.org/wiki/Extension:Math}{\mbox{MediaWiki}:~Math}
and MathSearch \cite{schubotz} extensions.
Within the DRMF, we will develop technology for users to
interact with formulae using a clipboard, which allows for easy copy/paste of 
formula source representations (to include \LaTeX\,with DLMF macros; presentation or content MathML; as well as input formats for 
computer algebra systems such as 
\href{http://www.wolfram.com/mathematica/}{Mathematica},
\href{http://www.maplesoft.com}{Maple},
and \href{http://www.sagemath.org}{Sage}).

The DRMF treats formulae as first class objects, describing them in 
formula home pages that currently contain:
(1) a rendered description of the formula itself (required);
(2) bibliographic citation (required);
(3) open section for proofs (required);
(4) list of symbols used and links to their definitions corresponding to
the DLMF macros (required);
(5) open section for notes relevant to the formula
(e.g., formula name, if the formula is a generalization or
specialization of some other formula, growth or decay conditions, 
links to errata pages, etc.);
(6) open section for external links;
(7) substitutions with definitions required to understand the
formula; and
(8) constraints the formula must obey.
For each formula home page there is a corresponding talk page,
and we are incorporating a strategy for handling the insertion of formula errata.
A major resource in our ability to implement effective and precise OPSF search 
will be the use of the DLMF macros in building the 
\LaTeX\,\,source for OPSF formulae and related mathematical data.  \\
\indent Next, we present an overview of the seed resources, which we plan
to incorporate within DRMF. We have been given permission and are seeding 
the DRMF with data from the
NIST DLMF \cite{NIST}. 
We have also been
given permission to and are seeding \LaTeX\,\,formulae data from 
\cite{Koekoeketal} (KLS).
We will also incorporate Tom
Koornwinder's companion of recent arXiv published additions to KLS \cite{KoornwinderKLSadd}.
We have also been given permission to incorporate seed formula data from 
\cite{ErdelyiHTF,ErdelyiTIT} (BMP). Efforts to
upload BMP data, as well as any book data without existing \LaTeX\,\,source, will
prove extremely difficult, since this effort will rely
on the use of mathematical optical character recognition (OCR) software
such as \href{http://www.inftyreader.org}{InftyReader} to produce
\LaTeX\,\,source for these formulae. Mathematical OCR is still in its nascence and
this effort is currently under consideration for feasibility of use.
We are in communication with other authors and publishers to gain access and
permission for other proven sources of mathematical OPSF formulae such as
\cite{
AAR,
GaspRah,
Grad,
Ismail} and we are are excited about the prospect of seeding proof data
by Victor Moll and collaborators (see for instance \cite{Glasseretal}).
For \LaTeX~source where DLMF macros are not present (such as KLS), we are developing 
tools which automate DLMF macro replacements.  Seeding and generating symbol lists are 
accomplished by converting \LaTeX~source into Wikitext, in an automated fashion.          
We use {\tt Pywikibot} to automate the uploading of Wikitext pages to our demo site.
\setcounter{footnote}{0}

\vspace{0.3cm}
\noindent {\bf\large Acknowledgements }\footnote{The mention of specific products, trademarks, or brand 
names is for purposes of identification only. Such mention is not to be interpreted in any way 
as an endorsement or certification of such products or brands by the National Institute of 
Standards and Technology, nor does it imply that the products so identified are necessarily 
the best available for the purpose. All trademarks mentioned herein belong to their 
respective owners.}\\[0.4cm]
\noindent We are deeply indebted 
to Deyan Ginev for sharing with us his expansive vision 
and especially for his support in the development of our proof of concept. Without 
his guidance and coding, our present demonstration would not be possible.  We would 
also like to thank Bruce Miller at NIST for his invaluable contributions 
regarding \LaTeX ML.  We would also like to express our deep gratitude to the 
KWARC group at Jacobs University, Bremen, Germany, and especially to its group 
leader, Michael Kohlhase, for his advice and for access to his group's mathweb server 
for our initial DRMF development.  We would also like to thank Dan Lozier, 
Tom Koornwinder, Dmitry Karp, Dan Zwillinger, Victor Moll, and Hans Volkmer for 
offering their advice and for valuable 
discussions.

\vspace{0.2cm}
\label{sect:bib}
\begingroup
\let\clearpage\relax


\endgroup

\end{document}